\documentclass[pr, 12pt, superscriptaddress]{revtex4}  
\usepackage{amssymb}
\usepackage{amsmath}
\usepackage{bbm}
\usepackage{graphicx}
\usepackage{color}

\begin{document}

\author{Philipp Werner}
\affiliation{Department of Physics, University of Fribourg, 1700 Fribourg, Switzerland}
\author{Sayed Ali Akbar Ghorashi}
\affiliation{Department of Physics and Astronomy, Stony Brook University, Stony Brook, NY 11794}

\title{Hidden Kondo lattice physics in single-orbital Hubbard models}

\date{\today}

\hyphenation{}

\begin{abstract} 
Single-orbital Hubbard models exhibit remarkably nontrivial correlation phenomena, even on nonfrustrated bipartite lattices. Some of these, like non-Fermi-liquid metal states, or the coexistence of heavy and light quasi-particles,  are reminiscent of the properties of more complex multi-orbital or Kondo-lattice systems. Here, we use basis transformations to map single-orbital models to effective multi-orbital descriptions and clarify how a ferromagnetic Kondo-lattice-like behavior emerges in prototypical models with flat bands or  van Hove singularities in the density of states: the Hubbard model on the diamond chain, square-lattice, Lieb lattice and honeycomb lattice. In particular, this mapping explains the non-Fermi-liquid states and pseudo-gaps found in the correlated metal regime. 
\end{abstract}

\maketitle


\section{Introduction}

Correlated electron systems exhibit some of the most remarkable phenomena found in condensed matter systems, including high-temperature superconductivity, colossal magnetoresistance, and correlation-driven metal-insulator transitions \cite{Imada1998,Dagotto2005}. The simplest model which captures correlation effects from local Coulomb repulsions is the single-orbital Hubbard model \cite{Gutzwiller1963,Hubbard1963}, which introduces an energy penalty $U$ for doubly occupied orbitals. This model describes the competition between localization and delocalization of electrons, and the Mott transition which occurs at half-filling, once $U$ exceeds a critical value comparable to the bandwidth. While exact solutions to the Hubbard model in dimensions $D=2$ and $3$ are not available, approximate numerical results indicate that the square lattice Hubbard model reproduces the phenomenology of cuprate superconductors \cite{Liechtenstein2000,Maier2005}, whose phase diagram in the doping-temperature plane exhibits a superconducting dome next to an antiferromagnetic Mott insulating phase. In the weakly doped regime, antiferromagnetic long-range order melts, giving rise to a pseudo-gapped metallic state with a momentum-dependent suppression of quasi-particles \cite{Kyung2006,Gull2009}, or a spin-glass phase \cite{Julien2003}. Above the superconducting dome, characteristic non-Fermi liquid properties, such as linear-$T$ resistivity \cite{Varma2020}, are observed. The spectral function of the weakly doped Mott phase reveals the coexistence of quasi-particles with strongly renormalized masses and more widely dispersing incoherent states \cite{Pruschke1993}. In the strongly overdoped regime, enhanced ferromagnetic correlations have been revealed in simulations \cite{Werner2020}, again in qualitative agreement with cuprate experiments \cite{Sonier2010,Kurashima2018}.

This remarkably complex physics of the single-orbital Hubbard model on the square lattice is in several respects reminiscent of the properties of more complex multi-orbital \cite{Biermann2005,Hoshino2016} or Kondo-lattice systems \cite{Yunoki1998,Dagotto1998}. The dominant magnetic ordering tendency in multi-orbital Hubbard models on a bipartite lattice switches from antiferromagnetic order close to half-filling to ferromagnetic order in the more strongly doped Mott regime \cite{Hoshino2016}. Orbital-dependent bandwidths naturally result in light and heavy particles, or in orbital-selective Mott states \cite{Koga2004}. Multi-orbital Hubbard models with Hund coupling furthermore exhibit bad metal behavior and non-Fermi-liquid scalings in extended regions of their phase diagrams \cite{Werner2008,Georges2013}. These Hund-metal properties originate from the freezing of local moments in the metallic phase, which occurs if the system is  in proximity to a high-spin Mott phase \cite{Werner2008,Ishida2010}. The ferromagnetic Kondo lattice model exhibits a similar phenomenology \cite{Werner2006}. 
 
The purpose of this article is to clarify how effective multi-orbital or ferromagnetic Kondo-lattice physics emerges in different single-orbital Hubbard models, specifically in the case of bipartite lattices featuring flat bands or van Hove singularities in the noninteracting density of states. 
The article is organized as follows: Sec.~\ref{sec_basis} describes a basis transformation which allows to map a single-orbital Hubbard system to an effective two-orbital Hubbard model with Hund coupling. This mapping is then used in Sec.~\ref{sec_analysis} to analyze Hubbard models on different lattices. The Conclusions of the study are presented in Sec.~\ref{sec_conclusions}.

\section{Basis transformation}
\label{sec_basis}

Several previous theoretical studies used basis transformations to map single-orbital lattice models to effective cluster-orbital or multi-orbital problems. In the context of cuprate physics and cluster dynamical mean field theory calculations, such transformations were employed in Refs.~\cite{Ferrero2009,Liebsch2009, Liebsch2011,Shinaoka2015,Werner2016}, and also the bandstructure of the diamond chain can be intuitively understood by introducing cluster orbitals \cite{Kobayashi2016}. 
Here, we consider the single-orbital Hubbard model with (usually negative) nearest neighbor hopping $t$ and on-site repulsion $U$,
\begin{equation}
H=t\sum_{\langle i,j\rangle\sigma} d^\dagger_{i\sigma} d_{j\sigma}+U\sum_i n_{i\uparrow}n_{i\downarrow},
\end{equation}
where $d_{j\sigma}$ denotes the annihilation operator for spin $\sigma$ on site $i$ and $n_{i\sigma}=d^\dagger_{i\sigma}d_{i\sigma}$. 
A simple and useful transformation is the mapping to bonding and antibonding orbitals,
\begin{equation}
c_\sigma=\frac{1}{\sqrt{2}}(d_{i\sigma}+d_{j\sigma}), \quad f_\sigma=\frac{1}{\sqrt{2}}(d_{i\sigma}-d_{j\sigma}),
\label{eq_basis}
\end{equation}
where $d_{i}$ and $d_{j}$ are two annihilation operators for 
different sites $i$ and $j$.
This transformation maps a hopping between the $d_i$ and $d_j$ orbitals onto a crystal field splitting,
\begin{equation}
t (d^\dagger_{i\sigma}d_{j\sigma}+\text{h.c.}) = \Delta_\text{cf}(n^c_\sigma-n^f_\sigma),
\label{eq_cf}
\end{equation}
with $\Delta_\text{cf}=t$ and $n^c=c^\dagger c$, $n^f=f^\dagger f$ the density operators for the $c$ and $f$ orbitals. For $t<0$, the $c$ ($f$) orbitals are shifted down (up). In the case of a complex hopping parameter $te^{i\phi}$, which might be realized by Floquet driving \cite{Kitagawa2011}, the energy splitting is given by the real part of the hopping, while the imaginary part induces an imaginary $c$-$f$ hybridization.

A particularly interesting property of the transformation (\ref{eq_basis}) is that it maps the local Hubbard interaction $U$ to a Slater-Kanamori-type spin-rotation invariant two-orbital interaction $H^\text{sk}$ for the $c$ and $f$ orbitals \cite{Shinaoka2015,Werner2016},
\begin{align}
U(n^d_{i\uparrow}n^d_{i\downarrow}+n^d_{j\uparrow}n^d_{j\downarrow})
&=\tilde U(n^c_{\uparrow}n^c_{\downarrow}+n^f_{\uparrow}n^f_{\downarrow})+\tilde U' \sum_\sigma n^c_{\sigma}n^f_{\bar\sigma}
-\tilde J(c^\dagger_{\downarrow}f^\dagger_{\uparrow}f_{\downarrow}c_{\uparrow} + f^\dagger_{\uparrow}f^\dagger_{\downarrow}c_{\uparrow}c_{\downarrow} + \text{h.c.})\nonumber\\
&\equiv H^\text{sk}_{sf}(\tilde U,\tilde U',\tilde J). 
\label{eq_sk}
\end{align} 
In the Slater-Kanamori expression, $\tilde U$ parametrizes the intra-orbital interaction, $\tilde U'$ the inter-orbital opposite-spin interaction, and $\tilde J$ the Hund coupling. We have $\tilde U=\tilde U'=\tilde J=U/2$, which implies that the inter-orbital same-spin interaction $\tilde U'-\tilde J'$ vanishes, while the Hund coupling is large ($\tilde J/\tilde U=1$). 

Because the ferromagnetic Hund coupling for $U>0$ favors aligned spins in the $c$ and $f$ orbitals, and hence in the two original $d$ orbitals, it is natural in a bipartite system to choose $d_i$ and $d_j$ on the same sublattice. In this case, the antiferromagnetic tendency of the Hubbard model on the bipartite lattice (for appropriate filling) translates into antiferromagnetic correlations between the high-spin composite moments. This makes the interpretation of the physics and the conceptual link to multi-orbital and Kondo-lattice descriptions transparent.     

An approximate mapping to a ferromagnetic Kondo lattice system becomes possible if the bandwidth or kinetic energy of the $f$ electrons is substantially smaller than that of the $c$ electrons, and if the $f$ orbitals are close to half-filled. We will show that this is indeed the case for relevant lattice models, like the diamond chain, square lattice, Lieb lattice and honeycomb lattice in the interesting filling regimes. This allows us to link phenomena like bad metal and non-Fermi-liquid behavior in the single-orbital Hubbard models to the spin-freezing in the ferromagnetic Kondo lattice model \cite{Werner2006}.

\section{Analysis of models}
\label{sec_analysis}

In this section, we apply the transformation (\ref{eq_basis}) to different Hubbard models on bipartite lattices to reveal the hidden Kondo lattice physics in these models. The local interaction of the Hubbard model is $U$ and the nearest neighbor hopping is denoted by $t$. We set the inter-unit cell spacing $a$ along the $x$ axis to 1.

\subsection{Diamond chain}
\label{sec_dimaond}

\begin{figure}[t]
\begin{center}
\includegraphics[angle=0, width=\columnwidth]{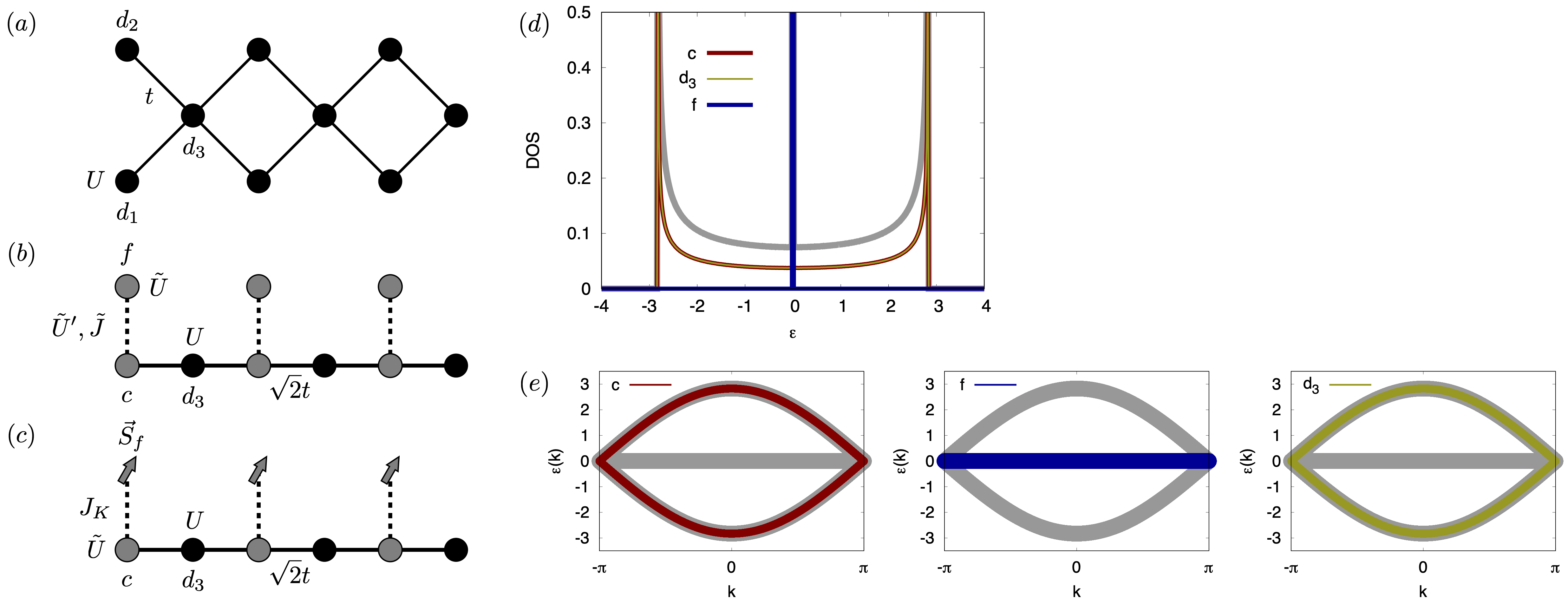}
\caption{(a) Diamond chain with hopping $t$ and interaction $U$. The unit cell has three sites with annihilation operators $d_1$, $d_2$, $d_3$. (b) Equivalent model after the bonding/antibonding transformation on $d_1$ and $d_2$.  The hopping between the $c$ and $d_3$ orbitals has amplitude $\sqrt{2}t$ and gray orbitals have an intra-orbital interaction $\tilde U=U/2$. The dashed line represents the inter-orbital terms of the Slater-Kanamori interaction $H^\text{sk}_{sf}$. (c) Approximate Kondo-lattice-type model for the half-filled system, with ferromagnetic Kondo coupling $J_K=-U$. (d,e) Partial densities of states and orbital characters of the bands in the noninteracting limit. The total density of states is shown by the gray line in (d). Energy is measured in units of $t$ and momentum in units of $1/a$.
}
\label{fig_diamond}
\end{center}
\end{figure}

The unit cell of the diamond chain contains three sites, with the original basis $d_1$, $d_2$, $d_3$, as indicated in Fig.~\ref{fig_diamond}(a). Performing the bonding/antibonding transformation on $d_1$ and $d_2$ decouples the $f$ orbital in the noninteracting system \cite{Kobayashi2016}, because the minus sign in the definition of the antibonding state (\ref{eq_basis}) results in cancellations of hopping terms.  The hopping Hamiltonian in the $\{c,d_3,f\}$ space,
\begin{equation}
H_\text{hop}(k_x,k_y)=\sqrt{2}t\left(
\begin{tabular}{ccc}
$0$ & $1+e^{ik}$ & $0$\\
$1+e^{-ik}$ & $0$ & $0$ \\
$0$ & $0$ & $0$
\end{tabular}
\right),
\end{equation}
thus features a flat $f$ band at energy $0$. In addition, the diagonalization yields two dispersing bands, which stem from the effective hopping with amplitude $\sqrt{2}t$ between the $c$ and the $d_3$ orbitals. The densities of states (DOS) and the orbital characters of the bands are shown in Fig.~\ref{fig_diamond}(d,e).

Adding the interactions, we obtain a chain with alternating $c$ and $d_3$ orbitals, in which the $c$ electrons are coupled by the Slater-Kanamori interaction $H_{cf}^\text{sk}(\tilde U,\tilde U',\tilde J)$ to the immobile $f$ electrons (Fig.~\ref{fig_diamond}(b)). At half-filling, we have on average a single electron in the $f$ orbitals, and in the strong interaction regime can replace these by a spin $\vec{S}_f=\frac{1}{2}f^\dagger_\gamma \vec{\sigma}_{\gamma\gamma'}f_\gamma$. This leads to a Kondo-lattice-like model, where the localized $f$ spins couple via a ferromagnetic $J_K=-U=-2\tilde J<0$ to the spins $\vec{S}_c=\frac{1}{2}c^\dagger_\gamma \vec{\sigma}_{\gamma\gamma'}c_\gamma$ of the $c$ electrons,
\begin{equation}
H_K=J_K \vec{S_f}\cdot \vec{S_c},
\end{equation}
see Fig.~\ref{fig_diamond}(c). The $d_3$ sites (with local interaction $U$) are not coupled to spins. 

A next-nearest neighbor hopping $t'$ between the $d_1$ and $d_2$ sites pushes the flat $f$ band up for $t'<0$ (Eq.~\eqref{eq_cf}) and opens a gap between the bands of $c$ and $d_3$ character. If the chemical potential is chosen such that the $f$ states are on average half filled, the effective Kondo lattice model still provides an approximate description of the interacting system.  

\begin{figure}[t]
\begin{center}
\includegraphics[angle=0, width=\columnwidth]{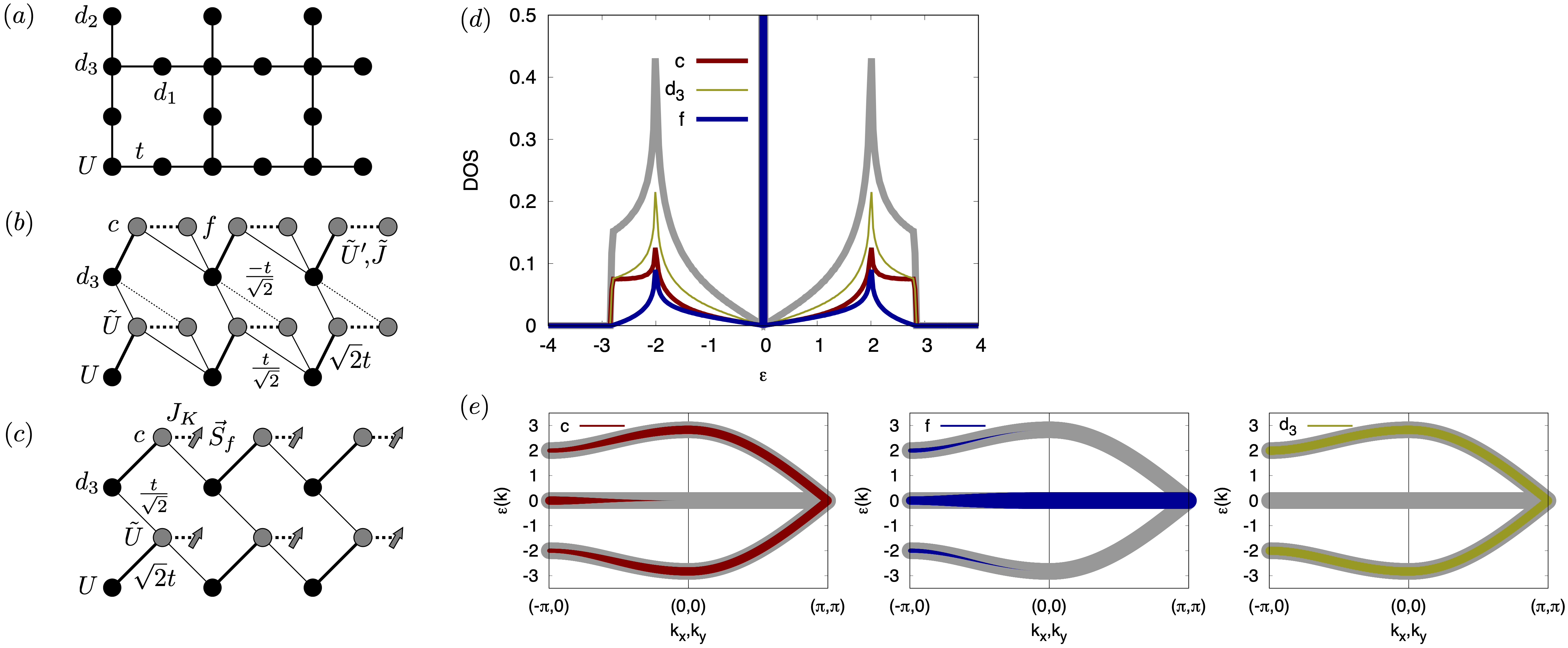}
\caption{(a) Lieb lattice with hopping $t$ and interaction $U$. The unit cell has three sites with annihilation operators $d_1$, $d_2$, $d_3$. (b) Equivalent model after the bonding/antibonding transformation on $d_1$ and $d_2$.  The hopping between the $c$ and $d_3$ orbitals has amplitude $\sqrt{2}t$ (within the unit cell) or $\frac{t}{\sqrt{2}}$ (between the unit cells), while the inter unit cell hopping between $f$ and $d_3$ is $\frac{t}{\sqrt{2}}$ (solid lines) or $-\frac{t}{\sqrt{2}}$ (dotted lines). Gray orbitals have an intra-orbital interaction $\tilde U=U/2$, and the dashed line represents the inter-orbital terms of the Slater-Kanamori interaction $H^\text{sk}_{sf}$. (c) Approximate Kondo-lattice-type model for the half-filled system, with ferromagnetic Kondo coupling $J_K=-U$. (d,e) Densities of states and orbital characters of the bands in the noninteracting limit. The total density of states is shown by the gray line in (d). Energy is measured in units of $t$ and momentum in units of $1/a$.
}
\label{fig_lieb}
\end{center}
\end{figure} 

\subsection{Lieb lattice}
\label{sec_lieb}

The two-dimensional Lieb lattice is similar to the diamond chain in the sense that it features two types of sites, with coordination number 4 and 2, respectively, and a three-site unit cell. Also in this case, the unequal number of sites in the two sublattices implies the existence of a flat band \cite{Lieb1989}. Denoting the annihilation operators for the sites in the unit cell by $d_1$, $d_2$, $d_3$ (Fig.~\ref{fig_lieb}(a)) and performing the bonding/antibonding transformation on $d_1$ and $d_2$ leads to the following hopping Hamiltonian  in the $\{c,d_3,f\}$ space,
\begin{equation}
H_\text{hop}(k_x,k_y)=\frac{t}{\sqrt{2}}\left(
\begin{tabular}{ccc}
$0$ & $2+e^{ik_x}+e^{ik_y}$ & $0$\\
$2+e^{-ik_x}+e^{-ik_y}$ & $0$ & $e^{-ik_x}-e^{-ik_y}$ \\
$0$ & $e^{ik_x}-e^{ik_y}$ & $0$
\end{tabular}
\right).
\end{equation}
In contrast to the diamond chain, the $f$ states are now not fully decoupled for $U=0$, but hybridized with the $d_3$ states, and the flat band has contributions both from the $f$ and $c$ states. The $f$ weight in the delta function contribution to the DOS (Fig.~\ref{fig_lieb}(d)) is however more than twice larger than the $c$ weight, and also the fat band plots in Fig.~\ref{fig_lieb}(e) show the dominant $f$ character of the flat band. Defining the effective half bandwidths $W_c$, $W_d$ and $W_f$ for the $c$, $d_3$ and $f$ electrons as the square root of the variance of the respective partial DOSes, we find $W_c=1.73t$, $W_d=2t$, $W_f=t$.  

For $U>0$, the Slater-Kanamori interaction $H^\text{sk}_{cf}$ acts on the $c$ and $f$ electrons, while the on-site repulsion for the $d_3$ electrons remains $U$ (Fig.~\ref{fig_lieb}(b)). At half-filling, given the orbital dependent bandwidths and large $J/U$ ratio, we can expect strongly orbital-dependent correlation effects, or even an orbital-selective Mott state for $U$ comparable to the bandwidth \cite{Koga2004}. If the $f$ electrons are in a Mott state, we can neglect their hopping and replace them by a spin $S_f$. In this approximate description, the model becomes a ferromagnetic Kondo lattice model on a brickwall lattice, with modulated hoppings and interactions, and $S_f$ spins coupled to only one sublattice (Fig.~\ref{fig_lieb}(c)). Note that the van Hove singularities are shifted in the approximate model from $\epsilon=\pm 2t$ to $\epsilon=\pm \sqrt{2}t$, but this should not make a qualitative difference for the low-energy physics close to half-filling.

\subsection{Square lattice}
\label{sec_square}

\subsubsection{Four-atom unit cell}

\begin{figure}[t]
\begin{center}
\includegraphics[angle=0, width=0.8\columnwidth]{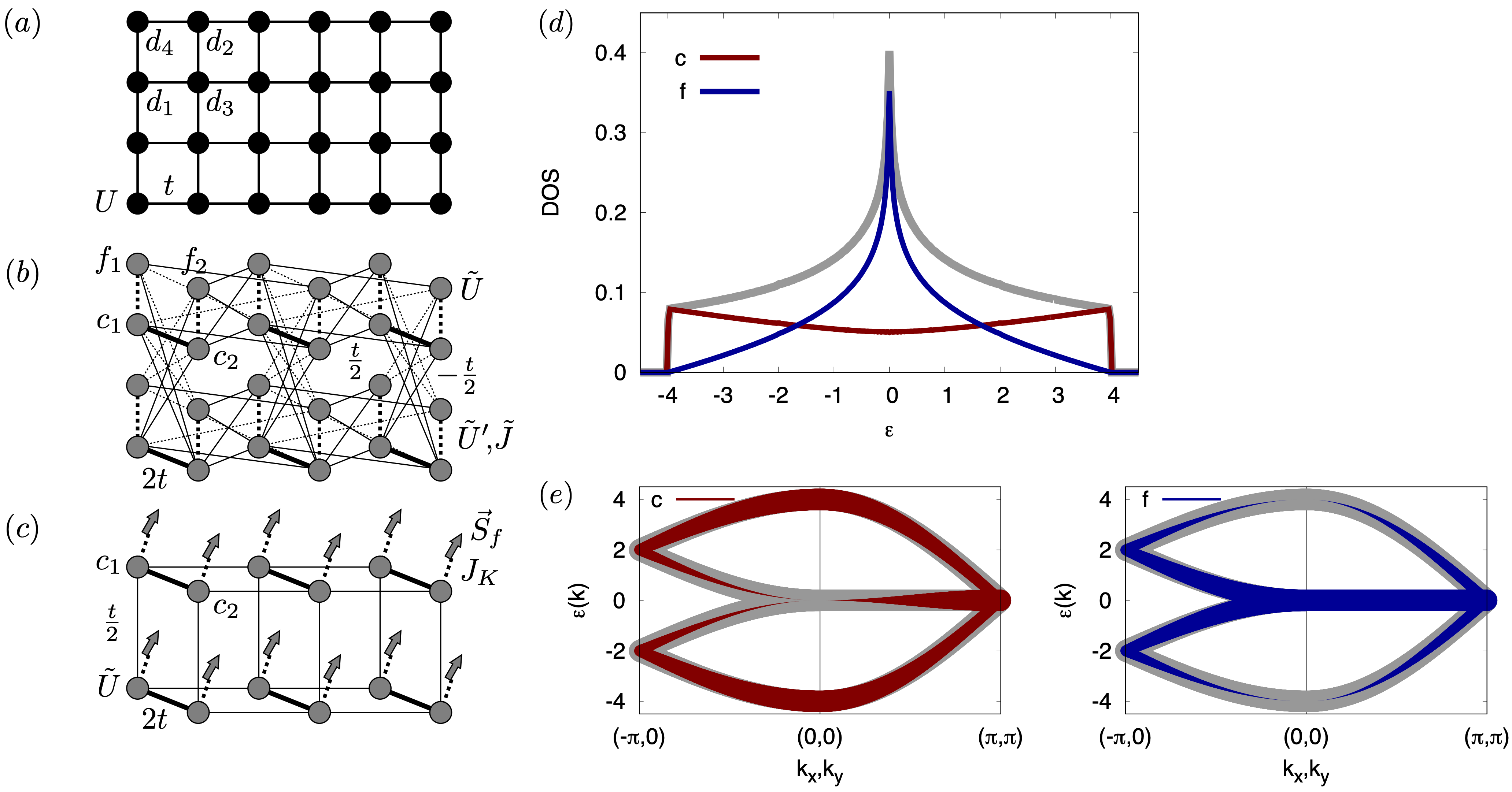}
\caption{(a) Square lattice with hopping $t$ and interaction $U$. We consider a unit cell of four sites with annihilation operators $d_1$, $d_2$, $d_3$, $d_4$. (b) Equivalent model after the bonding/antibonding transformation on $d_1$, $d_2$ and $d_3$, $d_4$, respectively, yielding the orbitals $c_1$, $f_1$ and $c_2$, $f_2$.  The hopping between the $c$ orbitals has amplitude $2t$ (within the unit cell) or $\frac{t}{2}$ (between the unit cells), while the inter unit cell hopping involving $f$ orbitals has amplitude $\frac{t}{2}$ (solid lines) or $-\frac{t}{2}$ (dotted lines). Gray orbitals have an intra-orbital interaction $\tilde U=U/2$, and the dashed line represents the inter-orbital terms of the Slater-Kanamori interaction $H^\text{sk}_{sf}$. (c) Approximate Kondo-lattice-type model for the system close to half-filling, with ferromagnetic Kondo coupling $J_K=-U$. (d,e) Densities of states and orbital characters of the bands in the noninteracting limit. The total density of states is shown by the gray line in (d). Energy is measured in units of $t$ and momentum in units of $1/a$.
}
\label{fig_square}
\end{center}
\end{figure}

In the case of the square lattice, we first consider a unit cell of four sites (a plaquette) \cite{Werner2016}, which maintains the $C_4$ symmetry (Fig.~\ref{fig_square}(a)). Performing the bonding/antibonding transformation along the diagonals of the plaquette ($d_1,d_2 \rightarrow c_1,f_1$ and $d_3,d_4 \rightarrow c_2,f_2$), we obtain a mapping to an effective two-orbital model with a complicated hopping structure (Fig.~\ref{fig_square}(b)). The on-site interaction $U$ in the original Hubbard model again translates into a Slater-Kanamori interaction $H^\text{sk}_{cf}$ for each pair of $c$ and $f$ orbitals. 

The hopping Hamiltonian written in the  $\{c_1,c_2,f_1,f_2\}$ space reads
\begin{align}\label{2D}
    &H_\text{hop}(k_x,k_y)=\left(\begin{array}{cc}
     A(k_x,k_y) & B(k_x,k_y)  \\
     B^\dagger(k_x,k_y)  & A(k_x,k_y)  
    \end{array}\right),\\
    &    A(k_x,k_y)=t\left(\begin{array}{cc}
     0 &  2+\cos(k_x)+\cos(k_y)  \\
     2+\cos(k_x)+\cos(k_y)   & 0 
    \end{array}\right),\\
     & B(k_x,k_y)=t\left(\begin{array}{cc}
     0 & -i(\sin(k_x)-\sin(k_y))  \\
     i(\sin(k_x)-\sin(k_y))  & 0 
    \end{array}\right),\\
     & C(k_x,k_y)=t\left(\begin{array}{cccc}
     0 & \cos(k_x)-\cos(k_y) \\
     \cos(k_x)-\cos(k_y) & 0      
    \end{array}\right).
\end{align}
While there is no flat band in this model, the noninteracting DOS of the square lattice exhibits a van Hove singularity at $\epsilon=0$. This singularity is contributed by the $f$ orbitals, while the partial $c$ DOS is rather flat, with the maximum weight near the band edges (Fig.~\ref{fig_square}(d,e)). Defining a $c$ and $f$ half-bandwidth from the square root of the variance of these partial DOSes, we obtain $W_c=2.45t$ and $W_f=1.41t$ \cite{Werner2016}. 

Given this difference in bandwidths, for a range of sufficiently large $U$ and near half-filling, we expect the $f$ electrons to be in a strongly correlated metallic or Mott insulating state with suppressed kinetic energy, while the $c$ electrons remain more weakly correlated. In this regime, we can simplify the model by eliminating the hoppings to the $f$ orbitals, which results in a bi-layer, square lattice ferromagnetic Kondo lattice model, with strong inter-layer hopping $2t$, intra-layer hopping $\frac{t}{2}$, on-site interaction $\tilde U$ for the $c$ electrons, and Kondo coupling $J_K=-U$ (Fig.~\ref{fig_square}(c)).  In this simplified model, the $c$-DOS features spurious van Hove singularities at $\epsilon=\pm 2t$, but it has the correct bandwidth of $8t$ and is featureless near $\epsilon=0$, where the flat $f$ band is located. The simplified model should thus qualitatively reproduce the physics near half-filling. 

\begin{figure}[t]
\begin{center}
\includegraphics[angle=0, width=0.8\columnwidth]{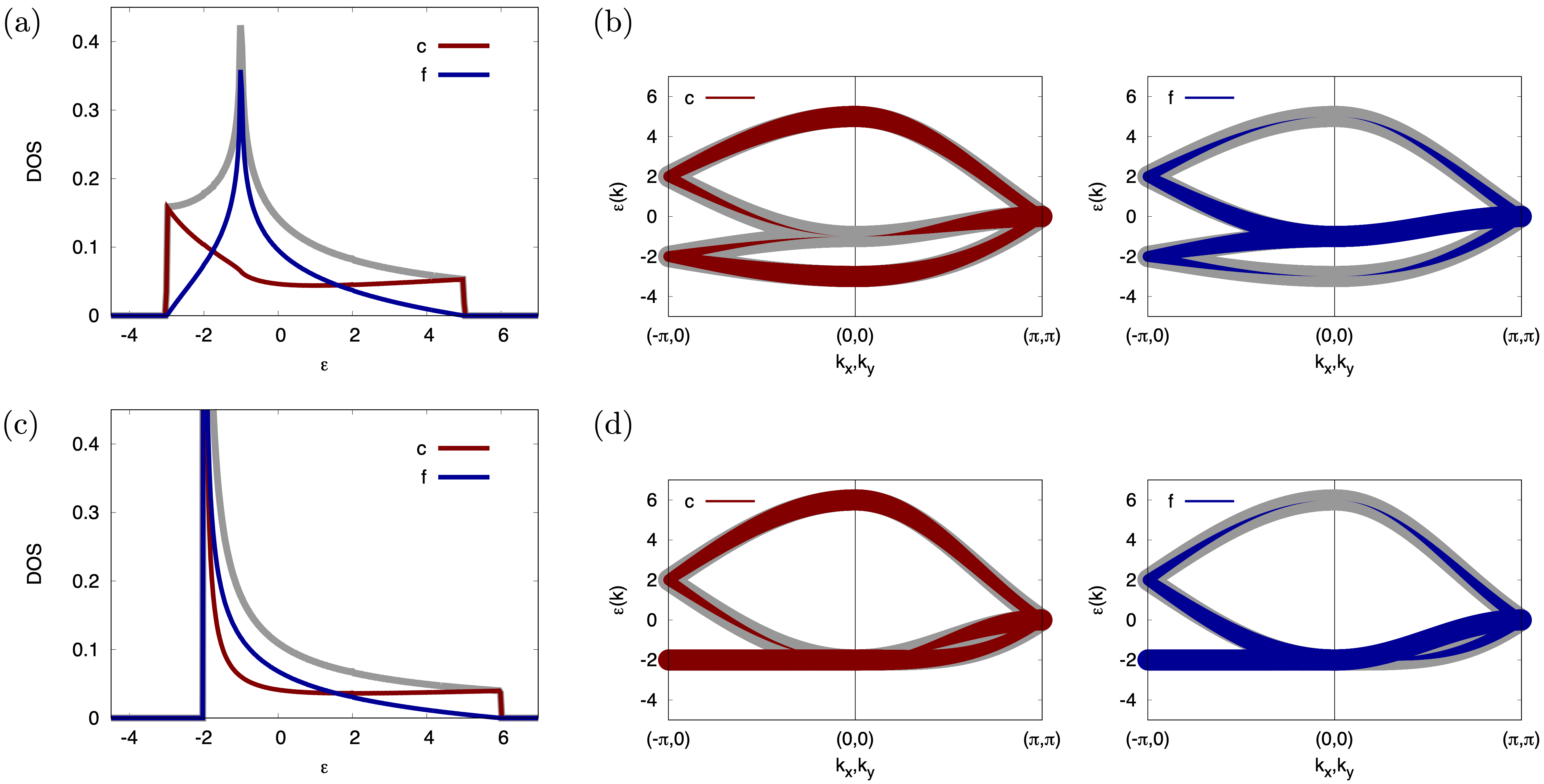}
\caption{Densities of states and orbital characters of the bands for the noninteracting square-lattice model with $t=-1$, $t'=-0.25t$ (a,b) and $t=-1$, $t'=-0.5t$ (c,d). The total DOS is shown by the gray lines in (a) and (c). Energy is measured in units of $|t|$ and momentum in units of $1/a$. 
}
\label{fig_square_tp}
\end{center}
\end{figure} 

In cuprate related studies, one often considers the Hubbard model with next-nearest neighbor hopping $t'$ (e. g. $t'=-0.3t$), and an interesting question is the nature of the non-Fermi liquid metal state that appears upon doping the Mott insulating system. If follows from Eq.~(\ref{eq_cf}) and the above bonding/antibonding transformation across the diagonals of the plaquette that the next-nearest neighbor hopping introduces a crystal field splitting between the $c$ and $f$ orbitals, which pushes the $f$ states down (for the usual choice of negative nearest-neighbor hopping $t<0$). The $t'$ term also adds new inter unit cell hoppings, so that the matrices $A$, $B$ and $C$ become
\begin{align}\label{2D}
& A=\left(\begin{array}{cc}
    t'+t'(\cos(k_x)+\cos(k_y))+t'\cos(k_x+k_y)  &  2t+t(\cos(k_x)+\cos(k_y))\\
     2t+t(\cos(k_x)+\cos(k_y))   & t'+t'(\cos(k_x)+\cos(k_y))+t'\cos(k_x-k_y) 
    \end{array}\right),\\
     & B=\left(\begin{array}{cc}
     it'(\sin(k_x)+\sin(k_y)) + i t' \sin{(k_x+k_y)}  &  -it(\sin(k_x)-\sin(k_y))\\
       it(\sin(k_x)+\sin(k_y))  & -it'(\sin(k_x)-\sin(k_y)) - it' \sin(k_x-k_y)
    \end{array}\right),\\
     & C=\left(\begin{array}{cc}   
     -t'-t'(\cos(k_x)+\cos(k_y))-t'\cos(k_x+k_y) &  t(\cos(k_x)-\cos(k_y))\\
     t(\cos(k_x)-\cos(k_y))   & -t'-t'(\cos(k_x)+\cos(k_y))-t'\cos(k_x-k_y)
    \end{array}\right).
\end{align}
Still, for realistic $t'$ the van Hove singularity is contributed by the $f$ electrons and the $c$ contribution to the DOS remains broad (Fig.~\ref{fig_square_tp}(a,b)). From studies of correlated multiorbital Hubbard models with crystal field splitting it is known that near half-filling, these tend to reshuffle the electrons between the orbitals in such a way that the most strongly correlated orbital remains half-filled \cite{Werner2009}. Hence, if the chemical potential is near the van Hove point, we expect an orbital selective Mott state or at least strongly enhanced correlation effects in the $f$ orbitals. The mapping to an effective multi-orbital model thus explains the coexistence of weakly renormalized and almost localized (at high temperature incoherent) electrons in the doped Mott insulating Hubbard model on the square lattice. In particular, the effective Kondo lattice description is applicable to the doped Mott regime and provides an intuitive explanation for the high-temperature bad metal state, where strong scattering off (frozen) composite magnetic moments destroys the quasi-particles \cite{Werner2006,Werner2016}. 

For $t'=-0.5t$, the dispersion of the Hubbard model becomes partly flat \cite{Sayyad2020}, leading to a strong peak at the lower edge of the DOS (Fig.~\ref{fig_square_tp}(c,d)). This peak has significant $c$ and $f$ contributions, and the chosen representation is no longer appropriate for classifying the electrons into almost localized and dispersive flavors.

\subsubsection{Momentum space picture}
\label{sec_momentum}

The physics of the square lattice Hubbard model in the doped Mott regime has been studied intensively with the dynamical cluster approximation (DCA) \cite{Hettler1998} and related schemes \cite{Maier2005}. In DCA calculations, the Brillouin zone is decomposed into patches and the self-energy is assumed to be constant in each patch. In DCA calculations, doping the Mott state induces a sector-selective Mott transition, where the patches at or near the nodes $(k_x,k_y)=(\pm \frac{\pi}{2},\pm \frac{\pi}{2})$ become metallic, while those at or near the antinodes $(k_x,k_y)=(\pi,0)$ and $(0,\pi)$ remain insulating \cite{Ferrero2009,Werner2009b}. Some studies have drawn an analogy to orbital selective Mott physics in multi-orbital Hubbard models, by calling this phenomenon an orbital-selective Mott transition in momentum space \cite{Ferrero2009,Gull2009}. This analogy is interesting, but it is difficult to make it precise because the Hubbard interaction expressed in momentum space has terms which are not present in a standard multi-orbital Hubbard interaction. 

\begin{figure}[t]
\begin{center}
\includegraphics[angle=0, width=0.8\columnwidth]{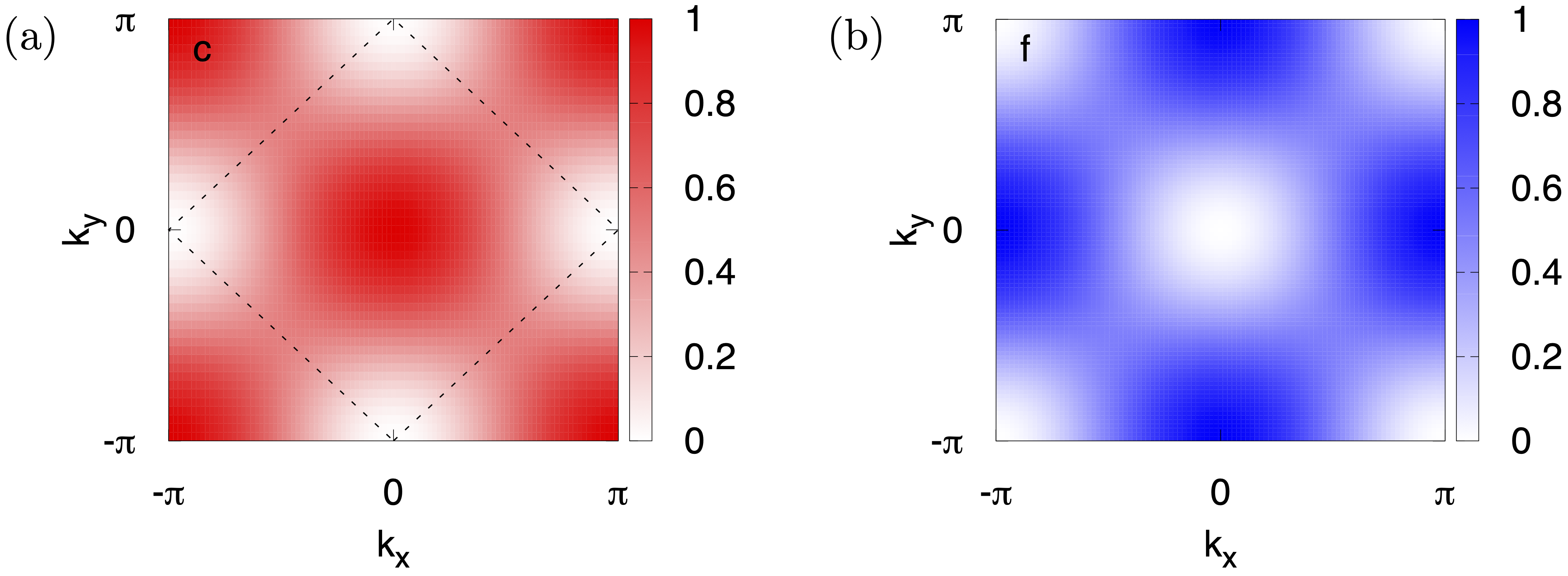}
\caption{Orbital character of the momentum states in the first Brillouin zone of the 2D square lattice. Panel (a) shows the $c$ orbital character and panel (b) the $f$ orbital character. The dashed line in (a) is the Fermi surface of the half-filled noninteracting system with $t'=0$.  
}
\label{fig_square_bz}
\end{center}
\end{figure}

In the context of this discussion it is however relevant to ask how the real-space bonding/antibonding transformation (\ref{eq_basis}) decomposes the first Brillouin zone of the square lattice model into momentum regions with $c$ and $f$ character. The transformation which maps the original site representation (operators $d_i$) to the diagonal band representation (operators $d_k$) is the Fourier transformation, in which we can split the sum over sites into a sum over plaquettes $I$ and an internal sum over the sites of each plaquette (index $j$):
%
$d_k=\frac{1}{\sqrt{N_I}}\sum_I\frac{1}{\sqrt{4}}\sum_j e^{i{\vec k}\cdot ({\vec R}_I+{\vec r}_j)} d_{I+j}$.
%
On each plaquette, we can apply the transformation (\ref{eq_basis}), which yields $d_{j=(0,0)}=\frac{1}{\sqrt{2}}(c_1+f_1)$, $d_{j=(1,1)}=\frac{1}{\sqrt{2}}(c_1-f_1)$, $d_{j=(1,0)}=\frac{1}{\sqrt{2}}(c_2+f_2)$, and $d_{j=(0,1)}=\frac{1}{\sqrt{2}}(c_2-f_2)$. The transformation from the $c/f$ to the momentum basis is thus
\begin{align}
d_k=\frac{1}{\sqrt{N_I}}\sum_I e^{i{\vec k}\cdot {\vec R}_I}\frac{1}{\sqrt{8}}&\Big[(e^{i{\vec k}\cdot (0,0)}+e^{i{\vec k}\cdot (0,0)}) c^{(I)}_1 
+ (e^{i{\vec k}\cdot (1,0)}+e^{i{\vec k}\cdot (0,1)}) c^{(I)}_2\nonumber\\
 &+(e^{i{\vec k}\cdot (0,0)}-e^{i{\vec k}\cdot (0,0)}) f^{(I)}_1 
+ (e^{i{\vec k}\cdot (1,0)}-e^{i{\vec k}\cdot (0,1)}) f^{(I)}_2\Big].
\end{align}
The $c$ and $f$ weight of the momentum point $k$ is given by the sum of the squared norms of the corresponding prefactors, which gives 
\begin{align}
w_c(k)=\frac{1}{2}(1+\cos(k_x)\cos(k_y)), \quad w_f(k)=\frac{1}{2}(1-\cos(k_x)\cos(k_y)).
\end{align}
These weights are plotted in Fig.~\ref{fig_square_bz} and are consistent with the partial DOSes (Fig.~\ref{fig_square}(a)), which imply that the saddle-point regions near the anti-nodes are $f$-like, while the band edges near $k=(0,0)$ and $k=(\pi,\pi)$ are $c$ like. 

In or near the orbital selective Mott regime, the $f$ electrons are Mott insulating or highly incoherent, while the $c$ electrons are metallic with well-defined quasi-particles. The Fermi surface (dashed line in Fig.~\ref{fig_square_bz}(a)) thus breaks up into segments (or arcs in the case of $t' \ne 0$), because only the nodal region of the noninteracting Fermi surface has a partial $c$ character and can host quasi-particles. The momentum space picture of our effective multi-orbital representation therefore provides an intuitive understanding for this much-debated phenomenon. 

The decomposition of the Brillouin zone into $c$ and $f$ dominated regions does not depend on the next-nearest neighbor hopping $t'$, so that the above discussion remains meaningful for not too large $t'$. As mentioned in the previous subsection, the $c/f$ representation is no longer particularly suitable for $t'=0.5t$, where the flat parts of the dispersion are not anymore confined to the antinodal region.

\begin{figure}[t]
\begin{center}
\includegraphics[angle=0, width=0.8\columnwidth]{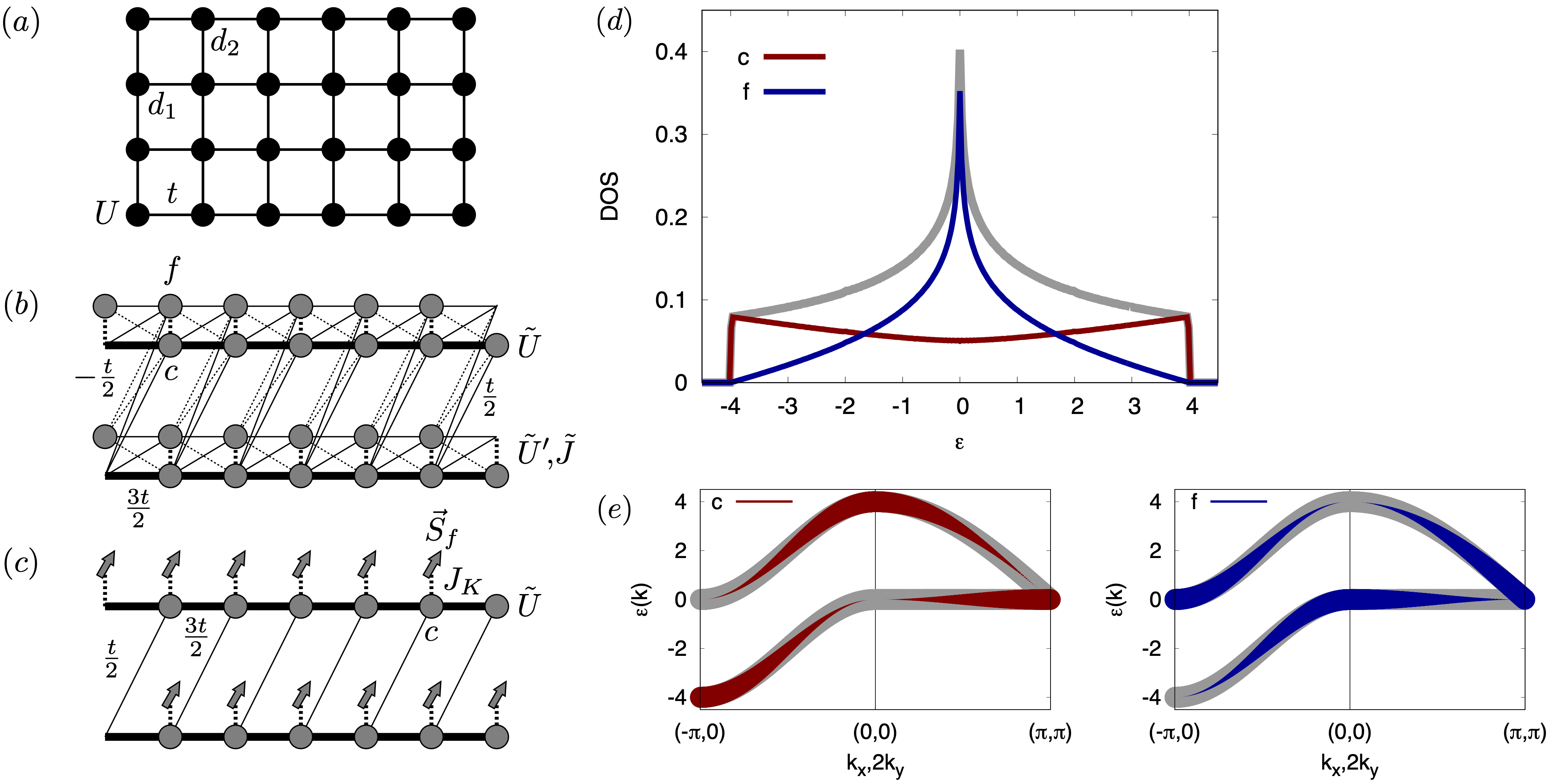}
\caption{(a) Square lattice with hopping $t$ and interaction $U$. We consider a unit cell of two sites with annihilation operators $d_1$ and $d_2$. (b) Equivalent model after the bonding/antibonding transformation on $d_1$, $d_2$, yielding the orbitals $c_1$, $f_1$.  The hopping between the $c$ orbitals has amplitude $\frac{3t}{2}$ (thick solid lines) or $\frac{t}{2}$ (thin solid lines), while the inter unit cell hopping involving $f$ orbitals has amplitude $\frac{t}{2}$ or $-\frac{t}{2}$ (dotted lines). Gray orbitals have an intra-orbital interaction $\tilde U=U/2$, and the dashed line represents the inter-orbital terms of the Slater-Kanamori interaction $H^\text{sk}_{sf}$. (c) Approximate Kondo-lattice-type model for the system close to half-filling, with ferromagnetic Kondo coupling $J_K=-U$. (d,e) Densities of states and orbital characters of the bands in the noninteracting limit. The total density of states is shown by the gray line in (d). Energy is measured in units of $t$ and momentum in units of $1/a_x$. 
}
\label{fig_square_simplified}
\end{center}
\end{figure} 

\subsubsection{Two-atom unit cell}
\label{sec_square_simplified}

We can also perform the bonding/antibonding transformation in a two-atom unit cell (with spacing $a_x=1$ and $a_y=2$), as indicated in Fig.~\ref{fig_square_simplified}(a). This choice breaks the $C_4$ rotation symmetry, but results in the same $c$ and $f$ electron densities of states as the four-atom unit cell (Fig.~\ref{fig_square_simplified}(e)). In particular, the $f$ electrons contribute the van Hove singularity at $\epsilon=0$. The Hamiltonian of the noninteracting model in the $\{c,f\}$ space reads
\begin{align}\label{2D}
    &H_\text{hop}(k_x,k_y)=t\left(\begin{array}{cc}
     3\cos(k_x)+\cos(k_x+2k_y) & i\sin(k_x)+i\sin(k_x+2k_y)  \\
     -i\sin(k_x)-i\sin(k_x+2k_y)  & \cos(k_x)-\cos(k_x+2k_y) 
    \end{array}\right),
\end{align}
and the corresponding hopping amplitudes are illustrated in Fig.~\ref{fig_square_simplified}(b). For $U>0$, the $c$ and $f$ electrons interact via the Slater-Kanamori interaction (\ref{eq_sk}).

Since the effective half-bandwidth of the $f$ electrons ($W_f=1.41t$) is significantly smaller than that of the $c$ electrons ($W_c=2.45t$), we may replace the $f$-electrons in the correlated regime near half-filling by spins, which results in the Kondo lattice model with anisotropic hoppings and $J_K=-U$ sketched in Fig.~\ref{fig_square_simplified}(c). The $c$-DOS of this simplified model is identical to that for the 4-site unit cell.

\subsection{Honeycomb lattice}
\label{sec_honeycomb}

\subsubsection{Four-atom unit cell}

As a fourth example, we discuss the honeycomb lattice. We are interested here in the behavior near 3/8 filling, where the chemical potential is in the vicinity of the van Hove singularity. In this region, correlation effects are enhanced, and various types of unconventional superconductivity have been theoretically proposed \cite{Xu2016}. For such doped systems, numerical simulations furthermore revealed a stripe order, which breaks the $C_6$ symmetry of the lattice \cite{Yang2021}.

\begin{figure}[t]
\begin{center}
\includegraphics[angle=0, width=0.8\columnwidth]{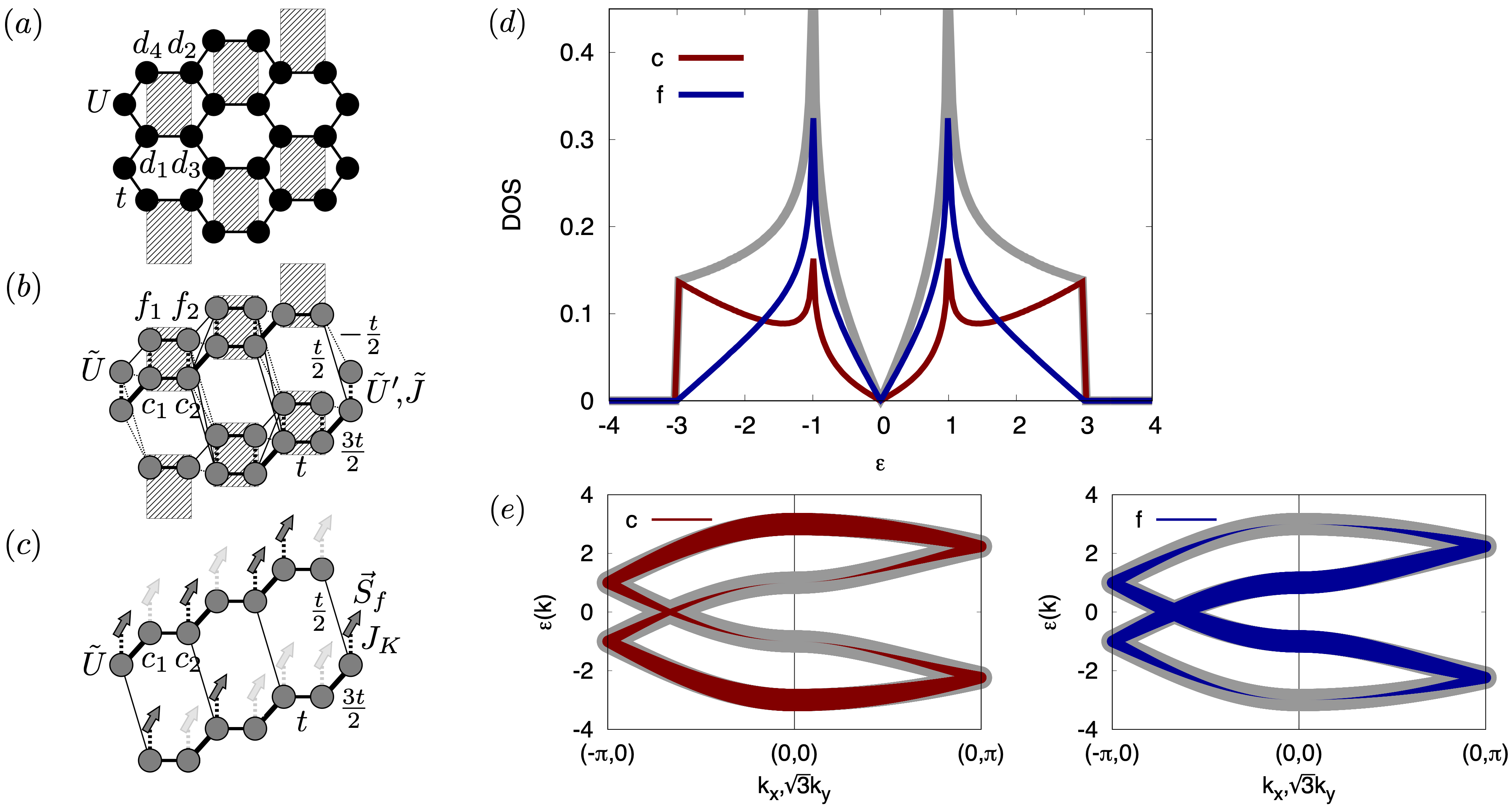}
\caption{(a) Honeycomb lattice with hopping $t$ and interaction $U$. We consider a unit cell of four sites (dashed rectangle) with annihilation operators $d_{1,2,3,4}$. (b) Equivalent model after the bonding/antibonding transformation on $d_1$, $d_2$ and $d_3$, $d_4$, respectively, yielding the orbitals $c_1$, $f_1$ and $c_2$, $f_2$.  The hopping between the $c$ orbitals has amplitude $\frac{3t}{2}$, $t$ or $\frac{t}{2}$ (thick, intermediate, and thin solid lines), while hoppings involving $f$ orbitals have amplitude $t$, $\frac{t}{2}$ or $-\frac{t}{2}$ (dotted lines). Gray orbitals have an intra-orbital interaction $\tilde U=U/2$, and the dashed line represents the inter-orbital terms of the Slater-Kanamori interaction $H^\text{sk}_{sf}$. (c) Approximate Kondo-lattice-type model for the system close to 3/8 filling, with ferromagnetic Kondo coupling $J_K=-U$ to a spin $S_f$ on half of the sites.  (d,e) Densities of states and orbital characters of the bands in the noninteracting limit. The total density of states is shown by the gray line in (d). Energy is measured in units of $t$ and momentum in units of $1/a_x$. 
}
\label{fig_honeycomb}
\end{center}
\end{figure}

The honeycomb lattice has a two-atom unit cell, but in order to perform the bonding/antibonding transformation within the same sublattice, we consider an extended four atom unit cell, as indicated by the hashed rectangles in Fig.~\ref{fig_honeycomb}(a). This choice is consistent with the striped solutions mentioned above. To define the superlattice vectors for this expanded unit cell, we rotate the $x,y$ coordinate system by 30 degrees relative to the horizontal axis. In this case, $a_x=1=\sqrt{3}l$ (with $l$ the bond length), and $a_y=3l$. After the transformation from $d_1$, $d_2$ to $c_1$, $f_1$ and $d_3$, $d_4$ to $c_2$, $f_2$, we obtain the following noninteracting Hamiltonian in the $\{c_1,c_2,f_1,f_2\}$ space:
\begin{align}\label{2D}
    &H_\text{hop}(k_x,k_y)=\left(\begin{array}{cc}
     A(k_x,k_y) & B(k_x,k_y)  \\
     B^\dagger(k_x,k_y)  & C(k_x,k_y)  
    \end{array}\right),\\
    &    A(k_x,k_y)=t\left(\begin{array}{cc}
     0 &  1+\frac{3}{2}e^{-ik_x} + \frac{1}{2}e^{i\sqrt{3}k_y}  \\
     1+\frac{3}{2}e^{ik_x} + \frac{1}{2}e^{-i\sqrt{3}k_y}   & 0 
    \end{array}\right),\\
     & B(k_x,k_y)=t\left(\begin{array}{cc}
     0 & -\frac{1}{2}e^{-ik_x} + \frac{1}{2}e^{i\sqrt{3}k_y} \\
     \frac{1}{2}e^{ik_x} - \frac{1}{2}e^{-i\sqrt{3}k_y} & 0  
    \end{array}\right),\\
     & C(k_x,k_y)=t\left(\begin{array}{cccc}
     0 & 1+\frac{1}{2}e^{-ik_x} - \frac{1}{2}e^{i\sqrt{3}k_y}  \\
     1+\frac{1}{2}e^{ik_x} - \frac{1}{2}e^{-i\sqrt{3}k_y}  & 0      
    \end{array}\right).
\end{align}

The density of states plot reveals that the van Hove singularities at $\epsilon=\pm t$ are mostly associated with the $f$ states, whose relative contribution to the DOS is more than twice larger in the van Hove region than that of the $c$ states. The latter, on the other hand, contribute most of the weight near the edges of the DOS at $\epsilon=\pm 3t$ (Fig.~\ref{fig_honeycomb}(d,e)). Focusing on the spectral weight at negative energies, and defining the effective half bandwidths of the $c$ and $f$ states as the square root of the second moment of the DOS relative to the van Hove point $\epsilon=-t$, we obtain $W_c=1.13t$ and $W_f=0.65t$. For a chemical potential near the van Hove point, and for $U>0$, the $f$ electrons thus exhibit stronger correlations than the $c$ electrons.  

In the noninteracting system, the $c$ electron filling at the van Hove point is $0.42$, while that of the $f$ electrons is $0.32$. Since interactions in two-orbital systems tend to shift the orbital occupations towards half- or quarter-filling \cite{Werner2009}, we expect that in the presence of $U$, the $f$ orbitals will be close to quarter-filled, while the $c$ orbitals will get close to half-filled. In a certain range of $U$ values, a metallic state with enhanced $f$ electron correlations or even an orbital selective Mott phase with insulating $f$ and metallic $c$ electrons will be realized.

The mapping to the effective two-orbital system yields the usual on-site interaction $\tilde U=U/2$ and the inter-orbital terms of the Slater-Kanamori type (Eq.~(\ref{eq_sk})), see Fig.~\ref{fig_honeycomb}(b). In the orbital-selective Mott state, or in the strongly correlated metal state near 3/8 filling,  the kinetic energy of the $f$ electrons is strongly suppressed, which motivates an approximate description in terms of a (depleted) ferromagnetic Kondo lattice model on a distorted honeycomb lattice, with exchange coupling $J_K=-U$, and bond-dependent hopping amplitudes for the $c$-electrons, as indicated in Fig.~\ref{fig_honeycomb}(c). In this model, assuming quarter-filled $f$ orbitals, only half of the sites couple to an $f$-electron spin, resulting in an annealed disorder problem with some similarity to the Falicov-Kimball model \cite{Falicov1969}. 

\begin{figure}[t]
\begin{center}
\includegraphics[angle=0, width=0.8\columnwidth]{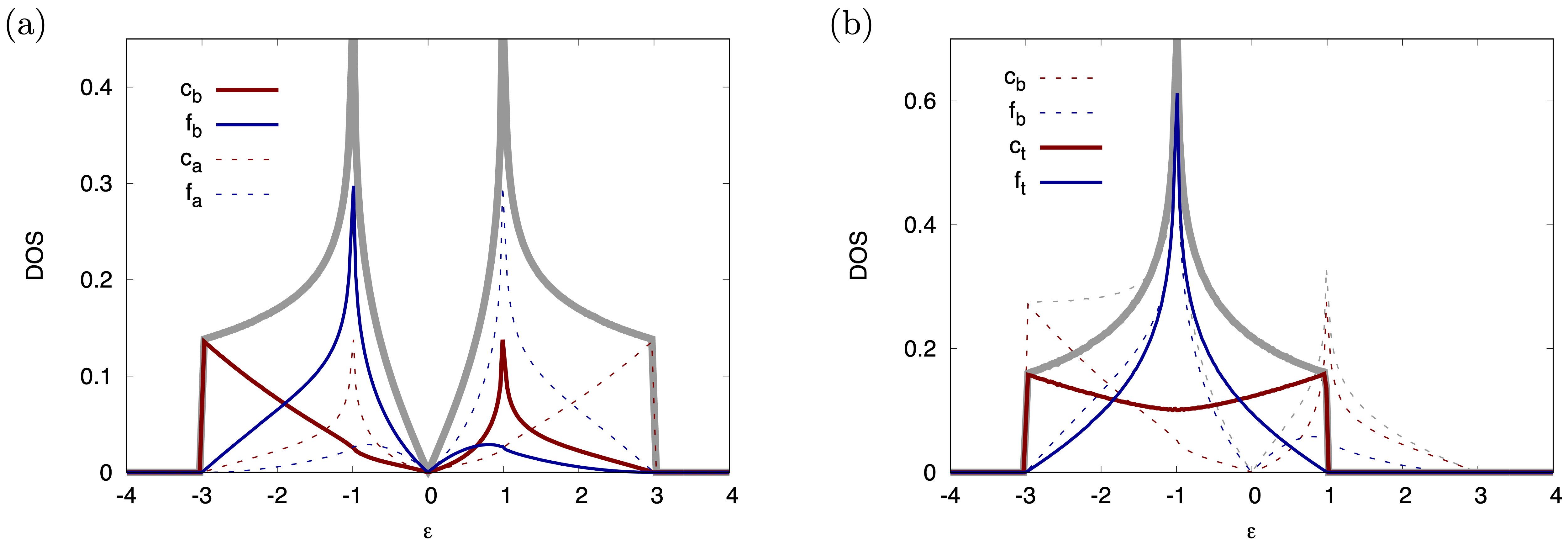}
\caption{(a) Partial DOSes for the bonding states $c_b$, $f_b$ (solid lines) and the antibonding states $c_a$, $f_a$ (dashed lines). (b) Partial DOSes for the simplified model $H_\text{bonding}^{c/f}$ (solid lines) and comparison to the partial $c_b$, $f_b$ DOSes (multiplied by a factor of two to match the integral). The total DOS is shown by the gray lines. We set $t=-1$ and measure energy in units of $|t|$. 
}
\label{fig_honeycomb_ab}
\end{center}
\end{figure} 

\subsubsection{Bonding $c/f$ states}

We may further simplify the description by switching to bonding and antibonding combinations of the $c/f$ orbitals and neglecting the (mostly empty for $t<0$) antibonding states. The transformation
\begin{align}
&c_b = \frac{1}{\sqrt{2}}(c_1+c_2), \quad c_a = \frac{1}{\sqrt{2}}(c_1-c_2),
&f_b = \frac{1}{\sqrt{2}}(f_1+f_2), \quad f_a = \frac{1}{\sqrt{2}}(f_1-f_2),
\end{align}
yields the partial densities of states illustrated in Fig.~\ref{fig_honeycomb_ab}(a) for the bonding and antibonding orbitals ($t=-1$). The $f_b$ orbital contributes the van Hove singularity, while the $c_b$ orbital dominates the states near the band edges. For a chemical potential at the van Hove point ($\mu=-1$), $58$\% of the noninteracting $f_b$ states and $63$\% of the $c_b$ states are occupied. The antibonding bands are on average $14$\% filled (consistent with a total filling of 3/8 per spin).

To simplify the description, we truncate the noninteracting Hamiltonian to the $\{c_b,f_b\}$ space, which yields
\begin{align}
& H_\text{bonding}^{c/f}(k_x,k_y)=t
\left(\begin{array}{cc}
     -1-\frac{3}{2} \cos(k_x) - \frac{1}{2} \cos(\sqrt{3} k_y) & -\frac{i}{2}(\sin(k_x)+\sin(\sqrt{3}k_y)) \\
     \frac{i}{2}(\sin(k_x)+\sin(\sqrt{3}k_y)) & -1-\frac{1}{2}\cos(k_x)+\frac{1}{2}\cos(\sqrt{3}k_y)
    \end{array}\right),
\end{align}
and the densities of states shown by the solid lines in Fig.~\ref{fig_honeycomb_ab}(b). We denote the orbitals of the truncated model by $c_t$ and $f_t$. After the truncation, the $c_t$ and $f_t$ DOS becomes symmetric around the van Hove point, resembling the result for the square lattice (Fig.~\ref{fig_square_simplified}). 

With interactions turned on, the bonding/antibonding transformation on the $c_{1,2}$ and $f_{1,2}$ orbitals generates a large number of terms: Slater-Kanamori type interactions between the bonding orbitals, between the antibonding orbitals, and between the bonding and antibonding orbitals of the same flavor, additional inter-flavor density-density interactions, as well as generalized spin-flip and pair hoppings which involve all four orbitals. The strength of all these interaction terms is $\hat U=U/4$. 

After elimination of the mostly empty antibonding states, we again end up with an effective two-orbital problem, with a Slater-Kanamori interaction (\ref{eq_sk}) with parameters $\tilde U=\tilde U'=\tilde J=\hat U$ between the $f_t$ and $c_t$ orbitals (note that this is a factor of two weaker than in the previous sections).

Assuming that $\hat U$ is strong compared to the effective width of the $f_t$ DOS, but comparable or lower than the effective width of the $c_t$ DOS, and for a chemical potential near the van Hove point, one expects a strongly reduced kinetic energy of the $f_t$ electrons, or even an orbital selective Mott phase with half-filled $f_t$ orbitals. We can then approximate the $f_t$ orbitals by spins $S_{f_t}$ and eliminate the hoppings to the $f_t$ states, resulting in a  $c$ electron dispersion of the form $H_\text{bonding}^{c}(k_x,k_y)=t(-1-\frac{3}{2} \cos(k_x) - \frac{1}{2} \cos(\sqrt{3} k_y))$. This corresponds to a Kondo lattice model on a square lattice with anisotropic hoppings $t_x=-\frac{3t}{4}$ and $t_y=-\frac{1}{4}$, similar to the effective model for the square lattice derived in Sec.~\ref{sec_square_simplified}, with the hoppings and interactions scaled down by a factor $1/2$. 

\section{Conclusions}
\label{sec_conclusions}

Multi-orbital and ferromagnetic Kondo-lattice physics emerges in various single-orbital Hubbard models with flat bands or van Hove singularities in the density of states. This can be understood by a simple transformation to bonding/antibonding combinations of pairs of sites (denoted by $f$ and $c$), which results in an exact two-orbital Hubbard model representation in which the flat parts or saddle points of the original dispersion are entirely or predominantly of $f$ character. Due to sign cancellations in the hopping contributions, the hybridization of the $f$ orbitals with the rest of the lattice vanishes, or is strongly suppressed, which leads to enhanced correlation effects in the $f$ orbitals. In a suitable interaction and filling range, the $f$ electrons become strongly correlated or Mott insulating, while the $c$ electrons remain moderately correlated and itinerant, which allows an approximate description of the system in terms of a ferromagnetic Kondo lattice with interacting $c$ electrons. The effective multi-orbital or Kondo lattice description naturally explains how heavy-Fermion like behavior and spin-freezing related non-Fermi liquid properties appear in single-orbital Hubbard systems. 

Our discussion also sheds light on similarities in the phase diagrams and physical properties of different classes of correlated materials. The ferromagnetic Kondo lattice model has been studied in connection with manganites \cite{Yunoki1998,Dagotto1998}, and it exhibits a rich phase diagram in the space of $c$ electron filling and $J_K$, including antiferromagnetic, ferromagnetic and phase separated regions. Long-range Coulomb interactions should however prevent the accumulation of charge in macroscopic regions, leading to charged stripes or other short range orderings \cite{Emery1993,Yunoki1998}. It is interesting that these results, which relate to the genuinely multiorbital manganites, resemble the physics of cuprates, whose essential properties are believed to be captured by a single-orbital Hubbard model. Cuprates are doped antiferromagnetic Mott insulators \cite{Imada1998}, which exhibit a tendency to some form of charge order at intermediate dopings \cite{Keimer2015} and enhanced ferromagnetic correlations at large doping \cite{Sonier2010,Kurashima2018}, all in qualitative agreement with the generic phase diagram of manganites. The bad metal and non-Fermi liquid behavior found in the underdoped and optimally doped regimes is furthermore reminiscent of the physics of Hund metals \cite{Werner2008,Haule2009,Georges2013}, e.~g. iron-based superconductors \cite{Hosono2015}, which are multi-orbital systems with ferromagnetic $J$. The mapping discussed here clarifies the link between these -- at first sight -- very different single-orbital and multi-orbital or Kondo lattice systems. 

\acknowledgements{We thank the Aspen Center for Physics for its hospitality during the summer 2023 program. P.W. acknowledges support from SNSF Grant No. 200021-196966 and S.A.A.G from the Air Force Office of Scientific Research under Grant No. FA9550-20-1-0260. P.W. thanks A. Georges and G. Kotliar for helpful feedback on the manuscript, and H. Aoki, S. Hoshino, H. Shinaoka and C. Yue for previous collaborations on topics related to this study.}

\end{document}